\documentclass[preprint,prc,preprintnumbers,superscriptaddress,amsmath,amssymb,floatfix]{revtex4-1}
\usepackage{graphicx,graphics,setspace,epsfig,color}
\usepackage[letterpaper,dvips,width=7.5in,height=8.5in,includemp=false]{geometry}
\usepackage[vcentermath]{}
\usepackage{epsf}
\usepackage{url}
\usepackage{amscd}
\usepackage{amsmath}
\usepackage[cm]{fullpage}
\usepackage{amsmath}
\usepackage{amssymb}
\usepackage{amsthm}
\usepackage{mathtools}
\usepackage{textcomp}
\usepackage{slashed}
\usepackage{graphicx}
\usepackage{url}
\usepackage{hyperref}
\usepackage[section]{placeins}
\usepackage{float}
\usepackage[active]{srcltx}
\usepackage{color}

\newcommand{\CL}{{\cal L}}

\long\def\symbolfootnote[#1]#2{\begingroup%
\def\thefootnote{\fnsymbol{footnote}}\footnote[#1]{#2}\endgroup}

\newcommand{\be}{\begin{equation}}
\newcommand{\ee}{\end{equation}}
\newcommand\beq{\begin{eqnarray}}
\newcommand\eeq{\end{eqnarray}} 

\newcommand{\E}{\mathcal{E}}

\usepackage{hyperref}
 
\begin{document}

\title{Charged-current reactions in the supernova neutrino-sphere}

\author{Ermal Rrapaj}
\email{ermal@uw.edu}
\affiliation{Department of Physics, University of Washington, Seattle, WA}
\affiliation{Institute for Nuclear Theory, University of Washington, Seattle, WA}
\author{J.\ W.\ Holt}
\affiliation{Department of Physics, University of Washington, Seattle, WA}
\author{Alexander Bartl}
\affiliation{Institut f\"ur Kernphysik,
Technische Universit\"at Darmstadt, 64289 Darmstadt, Germany}
\affiliation{ExtreMe Matter Institute EMMI,
GSI Helmholtzzentrum f\"ur Schwerionenforschung GmbH, 64291 Darmstadt, Germany}
\author{Sanjay Reddy}
\email{sareddy@u.washington.edu}
\affiliation{Institute for Nuclear Theory, University of Washington, Seattle, WA}
\affiliation{Department of Physics, University of Washington, Seattle, WA}
\author{A.\ Schwenk}
\affiliation{Institut f\"ur Kernphysik,
Technische Universit\"at Darmstadt, 64289 Darmstadt, Germany}
\affiliation{ExtreMe Matter Institute EMMI,
GSI Helmholtzzentrum f\"ur Schwerionenforschung GmbH, 64291 Darmstadt, Germany}

\begin{abstract}
We calculate neutrino absorption rates due to charged-current
reactions $\nu_e+n \rightarrow e^- + p $ and $\bar{\nu}_e+p
\rightarrow e^+ + n $ in the outer regions of a newly born neutron
star called the neutrino-sphere. To improve on recent work which has
shown that nuclear mean fields enhance the $\nu_e$ cross-section and
suppress the $\bar{\nu}_e$ cross-section, we employ realistic
nucleon-nucleon interactions that fit measured scattering phase
shifts. Using these interactions we calculate the momentum-, density-,
and temperature-dependent nucleon self-energies in the 
Hartree-Fock approximation.  A potential derived from chiral effective
field theory and a pseudo-potential constructed to reproduce
nucleon-nucleon phase shifts at the mean-field level are used to study 
the equilibrium proton fraction and
charged-current rates. We compare our results to
earlier calculations obtained using phenomenological mean-field models
and to those obtained in the virial expansion valid at low density and high temperature.
In the virial regime our results are consistent with previous calculations, 
and at higher densities relevant for the neutrino sphere, $\rho \gtrsim 10^{12}$ g/cm$^3$, we find the difference between the
$\nu_{e}$ and $\bar{\nu}_e$ absorption rates to be larger than
predicted earlier. Our results may have implications for
heavy-element nucleosynthesis in supernovae, and for supernova neutrino
detection.
\end{abstract}

\maketitle

\section{Introduction}

The neutrino opacity of dense matter plays a central role in
supernovae, associated nucleosynthesis, and the subsequent evolution of
the newly born neutron star called the proto-neutron star (PNS).
Neutrino interactions at the high densities and temperatures of relevance
are influenced by matter degeneracy, inter-particle correlations due
to strong and electromagnetic interactions, and by multi-particle
excitations~\cite{Reddy98,Burrows98,Burrows99,Reddy99,Hannestad98,%
Sedrakian:2000,Horowitz03,Lykasov08,Bacca08,Bacca11,Bartl:2014hoa}. Supernova
and PNS simulations that include these corrections have found them to play a
role in shaping the temporal and spectral aspects of neutrino
emission~\cite{Pons99,Reddy99,Huedepohl2010,Roberts12}. Of particular
interest to our study here are the spectra of electron and
anti-electron neutrinos, $\nu_e$ and $\bar{\nu}_e$, which decouple in the 
outer region of the PNS called the
neutrino-sphere. Here, the reactions $\nu_e+n \rightarrow e^- + p $
and $\bar{\nu}_e+p \rightarrow e^+ + n $ are an important source of
neutrino opacity, and their rates directly influence the mean energy of $\nu_e$
and $\bar{\nu}_e$ neutrinos~\cite{Reddy98,Burrows98,Burrows99,Reddy99}.
The mean neutrino energy can in turn impact supernova dynamics \cite{Janka12},
supernova nucleosynthesis~\cite{Qian96, Arcones2013}, and influence the number of
neutrinos detectable from a supernova in terrestrial neutrino
detectors~\cite{Scholberg2012}.

Since matter is neutron-rich in the neutrino-sphere, the reaction 
$\nu_e+n \rightarrow e^- + p $ is favored over $\bar{\nu}_e+p \rightarrow e^+ + n $, and on general grounds 
we can expect that  $\langle
\sigma_{\nu_e} \rangle > \langle \sigma_{\bar{\nu}_e}\rangle $, where
$\langle \sigma_{\nu_e} \rangle $ and $\langle
\sigma_{\bar{\nu}_e}\rangle$ are the thermally averaged neutrino and
anti-neutrino cross-sections, respectively. The
corresponding root-mean-square (rms) energies of neutrinos emerging
from the neutrino-sphere will satisfy the following inequality  $\epsilon_{\bar{\nu}_e} >
\epsilon_{\nu_e} $. It is now well established, through parametric
studies and simulations, that nucleosynthesis in the
neutrino-driven wind (NDW) is very sensitive to the difference $\delta
\epsilon=\epsilon_{\bar{\nu}_e} - \epsilon_{\nu_e}$. Neutron-rich
conditions in the material ejected by the neutrino-driven wind, a
prerequisite for the r-process, is only achieved when $ \delta
\epsilon > 4(m_n-m_p) \simeq 5$~MeV~\cite{Qian96, Arcones2013}. Parametric studies
indicate a robust r-process in the NDW is only
realized for an electron fraction $Y_e \lesssim 0.4$ which requires even larger $\delta \epsilon$~\cite{Hoffman97,Wanajo2013}. However,
recent simulations of supernova and PNS evolution do not achieve
these conditions, instead they predict $Y_e >0.45$~\cite{Huedepohl2010,Arcones2013,Fischer:2012}. This difficulty has led to a
renewed interest in charged-current reactions in the neutrino-sphere
to better determine the differences in neutrino spectra. 

The role of nuclear interactions in determining the charged-current rates in dense neutron-rich matter was first studied 
in \cite{Reddy98}. Subsequently, it was recognized \cite{Martinez-Pinedo12,Roberts12c} 
that the difference in the neutron and proton interaction energies 
enhances the electron neutrino absorption cross-section and simultaneously suppresses the cross-section for the
absorption of anti-electron neutrinos.  Simple phenomenological models based on the relativistic mean field (RMF) theory \cite{Roberts12c},  and a model independent approach based on the virial expansion valid at low density and high temperatures \cite{Horowitz12}  were used to calculate the difference between the neutron and proton interaction energies.  Using these inputs it was found that the electron neutrino absorption rate in the neutrino-sphere for typical thermal
neutrinos with energy $\simeq 10$ MeV could be enhanced by a factor of $2-4$, while the absorption rates for
anti-electron neutrinos were found to be suppressed by as much as an order of magnitude \cite{Martinez-Pinedo12,Roberts12c,Horowitz12}. 
Due to this suppression, other processes including the neutral-current processes such as
$ \bar{\nu}_e+\nu_e+ N + N \rightarrow N + N $ were found to play a role in
determining the $\bar{\nu}_e$ spectra, which were consequently found to be very
similar to the spectra expected for $\nu_\mu$ and
$\nu_\tau$.

In this article, we improve on these earlier studies by using realistic
nuclear interactions that can reproduce nucleon-nucleon (NN) phase
shifts to compute the nucleon self-energies and the equation of state
of hot and dense matter expected in the neutrino-sphere.  We use the
potential developed by Entem and Machleidt~\cite{Entem2003} within the
framework of chiral effective field theory (EFT) at
next-to--next-to--next-to--leading order (N$^3$LO) in the chiral
expansion. This low-momentum potential is able to reproduce low-energy
phase shifts without a strong repulsive core and it is expected that
many-body perturbation theory provides a reasonable description of
matter at moderate density and
temperature~\cite{Tolos:2007bh,Hebeler10,Hebeler:2010xb,Tews:2012fj,Gezerlis:2013ipa,Coraggio13,%
Holt2013,Kruger2013,Hagen2014,Drischler:2013iza,Coraggio14,Wellenhofer14}. To
assess the convergence of many-body perturbation theory in the
particle-particle channel (the ladder summation) for the partially degenerate conditions encountered in the neutrino-sphere, we define and use a 
pseudo-potential, which is given directly in terms of NN phase
shifts obtained from the partial-wave analysis (PWA) of the Nijmegen group~\cite{Stoks93}. The composition of matter, and the medium-induced self-energies
are obtained using finite-temperature perturbation theory in the
Hartree-Fock (HF) approximation.  This allows us to calculate the
in-medium Green's functions for neutrons and protons, and the 
density-, temperature-, and momentum-dependent nucleon dispersion
relations are naturally incorporated in calculations of the charged-current
cross-sections for $\nu_e$ and $\bar{\nu}_e$. We also present new results, using the formalism 
developed in Ref.~\cite{Bartl:2014hoa}, for the  neutrino pair absorption mean free path for the reaction 
$ \bar{\nu}_e+\nu_e+ N + N \rightarrow N + N $, which improves upon 
earlier work by properly accounting for nucleon-nucleon interactions and 
nucleon self-energies in the medium. 

In Sec.~\ref{sec:Kinematics} we describe the kinematics of
charged-current reactions and highlight the importance of nucleon
dispersion relations. The nucleon dispersion relation and the
composition of matter in the neutrino-sphere are calculated in
Sec.~\ref{sec:HFenergies}, where we also briefly discuss the NN
interactions used and assess the validity of the HF approximation for
the relevant conditions. In Sec.~\ref{sec:WeakRates} the neutrino-absorption 
rates using the HF self-energies are calculated and compared to results obtained in
earlier work. In
Sec.~\ref{sec:Conclusions} we discuss the implications of our findings
and identify areas where improvements are necessary. Finally, we note
that throughout we use natural units: we set $\hbar = 1$, the speed of
light $c=1$ and the Boltzmann constant $k_B=1$. Energy and temperature
are measured in MeV, and the density is measured in units of nucleons
per fm$^{3}$.

\section{Kinematics}
\label{sec:Kinematics}

\begin{figure}[h]
\centering
\includegraphics[scale=0.5]{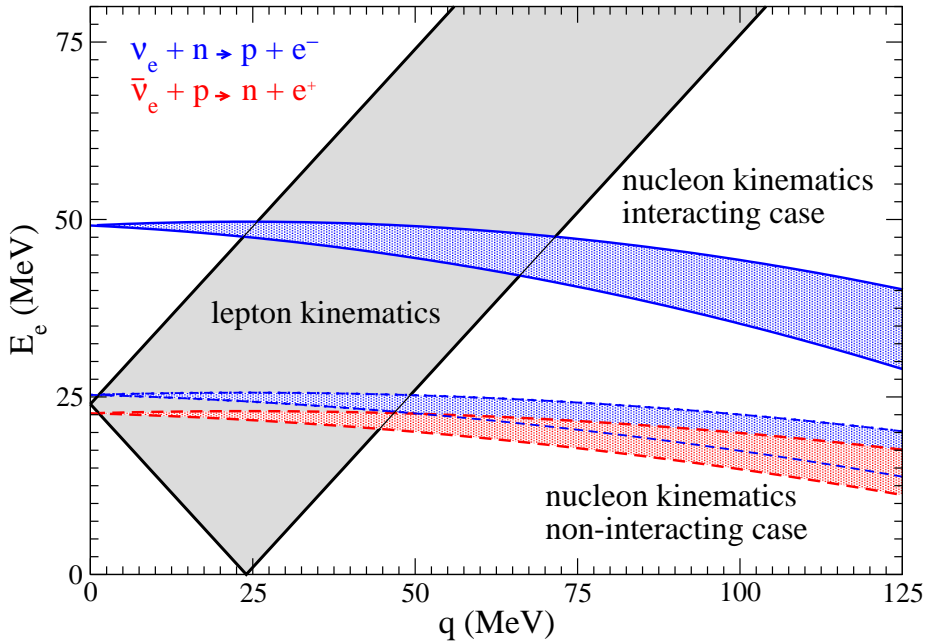}
\caption{(Color online) Energy and momentum constraints on the 
charged-current reactions for conditions discussed in the text.
Reactions are possible when the allowed
region for lepton kinematics, shown by the shaded region enclosed by
the black lines, overlaps the allowed region for nucleon kinematics,
shown by the regions enclosed by blue and red lines
corresponding to the $\nu_e$ and $\bar{\nu}_e$ reactions,
respectively. The region enclosed by the solid blue lines includes the nuclear 
self-energy difference for the transition $n \rightarrow p$  associated with the $\nu_e$ reaction, 
and regions enclosed by the dashed lines are for non-interacting nucleons. The $p \rightarrow n$ transition associated 
with the $\bar{\nu}_e$ reaction is kinematically forbidden as there is no overlap when nucleon self-energy corrections are included.}
\label{fig:kinematics}
\end{figure}

We begin with a general discussion of the kinematics of
charged-current reactions to highlight the importance of nuclear
interactions. Kinematic restrictions for the charged-current reactions
$\nu_e + n \rightarrow e^{-} + p$ and $\bar{\nu}_e + p \rightarrow
e^{+} + n$ are relevant, because the neutrino energy is comparable to
the typical energy and momentum scales in the hot and dense plasma in
the neutrino-sphere.  Due to strong electron degeneracy, final-state blocking
suppresses the $\nu_e$ absorption when the neutrino energy is
comparable or smaller than the electron Fermi energy. Similarly,
$\bar{\nu}_e$ absorption requires a neutrino energy large enough to
overcome the energy difference between the proton in the initial state
and the neutron plus positron energy in the final state. These
constraints are depicted in Fig.~\ref{fig:kinematics}, where we
illustrate energy and momentum conservation for an incoming neutrino
of energy $E_\nu=24$ MeV; this is the typical thermal energy of the neutrino for an ambient temperature of $T = 8$ MeV. The $x$-axis is the magnitude of the
momentum transferred to the nucleons, $\vec{q}=\vec{k}_\nu-\vec{k}_e$,
where $\vec{k}_\nu$ and $\vec{k}_e$ are the $\nu_e~(\bar{\nu}_e)$ and
final state $e^-~(e^+)$ lepton momenta, respectively. The $y$-axis is the
final-state lepton energy $E_e$. The shaded area enclosed
by the solid black lines is the region allowed by lepton kinematics
for an incoming neutrino with $E_\nu=24$ MeV.

The reaction can proceed when the allowed regions for nucleon and
lepton kinematics overlap.  Energy and momentum constraints imposed by
the nucleons for the $\nu_e + n \rightarrow e^{-} + p$ and
$\bar{\nu}_e + p \rightarrow e^{+} + n$ reactions are shown by the
regions enclosed by the dashed blue and red curves, respectively. For
the $\nu_e$ reaction, the blue region is defined by the equation
$E_n(|\vec{k}\,|)-E_p(|\vec{k}+\vec{q}\,|) =-\omega$, and for the
$\bar{\nu}_e$ reaction the red region is defined by
$E_p(|\vec{k}\,|)-E_n(|\vec{k}+\vec{q}\,|) =-\omega$, where $\omega =
E_\nu-E_e $ is the energy transferred to the nucleons. When nuclear interactions are neglected, 
the neutron and proton single-particle energies are given by $E_{n}(|\vec{k}\,|)=M_n+k^2/2M_n$ and
$E_{p}(|\vec{k}\,|)=M_p+k^2/2M_p$, respectively.  In this case, the allowed kinematic region for the $\nu_e$
and  $\bar{\nu}_e$ are similar and the small difference arises solely due to the small neutron-proton mass difference. 

In an interacting system, the single-particle energy is given by
\be 
E_{i=n,p}(|\vec{k}\,|)=M_i+\frac{k^2}{2M_i} + \Sigma_i(k) \equiv \varepsilon_i(k) + M_i\,,
\label{eq:n_dispersion}
\ee
where $\Sigma_i(k)$ is the momentum-, density-, and
temperature-dependent self-energy (we note that
in general, the self-energy will also be energy-dependent, but in the Hartree-Fock approximation employed
in the present study this does not arise).
At the densities $\rho \simeq 10^{11}-10^{13}$~g/cm$^3$ and
temperatures $T\simeq 3-10$~MeV of interest in the neutrino-sphere,
matter is very neutron-rich with an electron fraction $Y_e$ of only a
few percent (note that charge neutrality requires the
proton fraction $Y_p=Y_e$). Due to this large asymmetry, the proton and neutron self-energies are not equal,
$\Sigma_n(k) \ne \Sigma_p(k)$. Both neutron and proton energies are shifted downwards by the nuclear
interaction at the
densities and temperatures encountered in the neutrino-sphere, i.e., $\Sigma_i < 0$, because NN interactions are on
average attractive at the relevant low momenta ($k<200$~MeV). However, the 
energy shift is much larger for the protons and $\Sigma_n - \Sigma_p > 0$ because of the denser neutron background and the additional attraction in the neutron-proton interaction.    
This energy difference is related to the potential part of the nuclear symmetry energy --- in neutron-rich matter it costs nuclear interaction energy to convert protons to neutrons,
and there is an energy gain resulting from the conversion of neutrons to protons. The resulting change in the reaction $Q$ value modifies
the relative $\nu_e$ and $\bar{\nu}_e$ absorption rates as described below.

Using calculations of $\Sigma_n(k)$ and $\Sigma_p(k)$, which will be
discussed in detail in Sec.~\ref{sec:HFenergies}, we illustrate
the change in reaction kinematics in Fig.~\ref{fig:kinematics} by
enclosing the allowed nucleon kinematic regions (using the same color legend) by solid lines. The  $Q$ value for the reaction at $q=0$ is the energy shift $\Sigma_n(k)-\Sigma_p(k) \simeq
30$~MeV which  is much larger than the rest mass difference $M_n-M_p=1.3$ MeV.  This large energy gain associated
with $n \rightarrow p$ conversion shifts the outgoing electron energy
to larger values and the overlap region between lepton and nucleon kinematic regions
is enhanced. Further, the higher $Q$ value also helps overcome the 
Pauli blocking in the final state for  the degenerate electrons with $\mu_e/T \gtrsim 3-10$.  
In contrast, the $\bar{\nu}_e$ reaction is now kinematically forbidden because the $\bar{\nu}_e$ energy $E_\nu=24$ MeV
is insufficient to overcome the energy threshold $\simeq 30$ MeV to convert protons to neutrons. 

\section{Nucleon single-particle energies in the neutrino-sphere}
\label{sec:HFenergies}

\begin{figure}[h]
\centering
\includegraphics[scale=0.8]{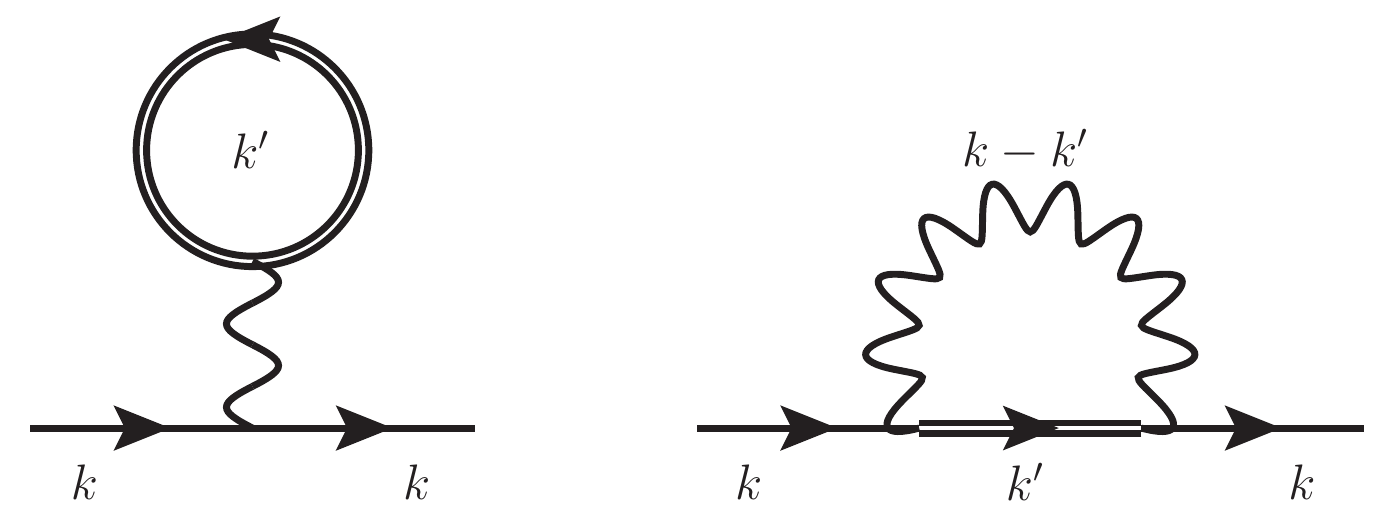}
\caption{Feynman diagrams depicting the self-consistent HF self-energy.
The double lines are the dressed nucleon propagators and wavy lines
represent the NN interaction.}
\label{fig:diagram}
\end{figure}

Nucleon dispersion relations are modified in a hot and dense
medium due to nuclear interactions.  In this section, we calculate
these modifications using realistic nuclear interactions in the HF
approximation. 
The self-consistent HF self-energy $\Sigma_{\rm HF}$ is defined
through the Feynman diagrams shown in Fig.~\ref{fig:diagram}.
We calculate $\Sigma_{\rm HF}$ using the finite-temperature
imaginary-time formalism and find the standard expression
\begin{equation}
\Sigma(\vec{k}) = \:\:\: \mathclap{\displaystyle\int}\mathclap{\textstyle\sum} \quad \frac{d^4k'}{(2 \pi)^4} \frac{\overline{V}
(\frac{\vec{k}-\vec{k'}}{2},\frac{\vec{k}-\vec{k'}}{2})}{i \nu_{k'}  - \xi(\vec k')} 
=\int\frac{d^3k'}{(2 \pi)^3}\, \overline{V}\! \left (\frac{\vec{k}-\vec{k'}}{2},\frac{\vec{k}-\vec{k'}}{2}\right ) f(\xi(\vec k')) \,,
\label{eq:HF}
\end{equation}
where $\xi(\vec k') = \varepsilon(\vec k')-\mu = {k'}^2/2M + \Sigma(k') - \mu$
is the single-particle energy measured with respect to the non-relativistic chemical
potential  (the rest mass of the nucleon has been subtracted). The sum over Matsubara frequencies is performed to
obtain the Fermi distribution function $f(\xi(k'))$. Since
the potential $\overline V$ is antisymmetrized, both contributions in
Fig.~\ref{fig:diagram} (the Hartree contribution on the left and the
Fock contribution on the right) are contained in the single expression
above. We use a spherical decomposition to represent the
anti-symmetrized potential in a partial-wave basis:
\begin{equation}
\langle \vec p Sm_sT| \overline{V} | \vec p^{\,\prime} Sm'_sT \rangle=(4 \pi)^2\! \sum_{l,m, l', m',J, M} \! i^{l'-l}\,Y_l^m(\hat{p}) Y_{l'}^{m'*}(\hat p^{\, \prime}) C^{J M}_{l m S m_s}C^{J M}_{l' m' S {m'_s}}\langle p|V_{l l' S}^{J T}|p'\rangle (1-(-1)^{l+S+T}) \,,
\label{eq:vsph}
\end{equation}
where $\vec p$ and $\vec p{\,'}$ are relative momenta and $\overline V
\equiv V (1-P_{12}) = V (1-(-1)^{l+S+T})$, with $P_{12}$ the
particle-exchange operator. The other symbols appearing in
Eq.~(\ref{eq:vsph}) have the standard meaning: $l,S, J$ and $T$ are
the relative orbital angular momentum, spin, total angular momentum
and total isospin quantum numbers of the nucleon pair, and the projections
of $\vec{S}$ and $\vec{l}$ onto the $z$-axis are given by the quantum
numbers $m_s$ and $m$, respectively.

For pure neutron matter the self-energy can be written as
\begin{equation}
\Sigma_{n}(\vec{k})=\frac{1}{2\pi} \int_0^{\infty} {k'}^2 dk' \int_{-1}^{1} d \cos \theta_{k'}\, f ( \xi(\vec{k'}) )
\sum_{l,S,J} (2J+1)\big \langle \, | (\vec
k - \vec {k'}\,)/2 |\, \big|\overline{V}^{J1}_{l l S} \big | \, | (\vec
k - \vec {k'}\,)/2 | \, \big \rangle \,,
\label{eq:neutron_only}
\end{equation}
where $\theta_{k'}$ is the angle between $\vec {k'}$ and $\vec
k$. The self-consistent solution to Eq.~(\ref{eq:neutron_only}) can be
obtained by iteration. To simplify notation we set $p=|\frac{1}{2} (\vec
k - \vec {k'}\,)|$ in the following. In asymmetric matter, containing
neutrons and protons we obtain the following coupled equations:
\begin{equation}
\Sigma_{m_t}(\vec{k})=\frac{1}{2\pi} \int_0^{\infty} {k'}^2 dk' \int_{-1}^{1} d \cos \theta_{k'} \, \sum_{l,S,J,T,m_t'} (2J+1)\, 
|{\cal C}_{\frac{1}{2}\, m_t \,\frac{1}{2}\, m_t'}^{T\, m_t+m_t'}|^2 \, \langle p |\overline{V}^{JT}_{l l S} | p \rangle \,
f(\varepsilon_{m_t'} ( \vec{k'} ) -\mu_{m_t'}),
\end{equation}
where $m_t$ and $m_t'$ label the isospin of the external and intermediate-state nucleon, respectively.

At low densities and high temperatures, where the
neutron fugacity satisfies $z_n = e^{\mu_n/T} \ll1$, the virial
expansion provides a model-independent
benchmark~\cite{Horowitz:2005nd,Horowitz2006}. This allows us to assess the 
validity of the HF approximation at densities characteristic of the neutrino-sphere.  First, we analyze the
HF predictions for the energy per particle in pure neutron matter in the
density range $n_B=0.001-0.02$~fm$^{-3}$ and temperature range
$T=5-10$ MeV. To calculate the energy
density in the HF approximation we use two approaches. In the first,
we employ the chiral N$^3$LO NN potential of Ref.~\cite{Entem2003},
and in the second approach we define and use the pseudo-potential. 

In the HF calculation, the
N$^3$LO potential is treated in the Born approximation. In contrast,
the pseudo-potential defined by the relation 
\be
\langle p | V^{pseudo}_{l l SJ}| p \rangle =
-\frac{\delta_{lSJ}(p)}{p M} \,,
\label{pseudo}
\ee
is constructed from the measured nucleon-nucleon phase shifts $\delta_{lSJ}(p)$ and 
 should be viewed as including a resummation of the ladder diagrams 
in the particle-particle channel. It is also known to correctly predict the energy shift in a system 
containing Fermions interacting strongly with a heavy impurity and is known in the context of 
condensed matter physics as Fumi's theorem \cite{Mahan}. In the following we show that the pseudo-potential  when used in the HF approximation 
reproduces the energy shift predicted by the virial expansion which is known to be exact in the limit of low density and high temperature. 

In the virial expansion, two-body interactions are included through the second
virial coefficient $b_2$, which is directly related to scattering phase shifts and is given by
\be
b_2 = \frac{1}{\pi \sqrt{2}\, T} \int_0^\infty d\epsilon \, e^{-\epsilon /2T} \, \sum_{lSJ} (2J+1) \delta_{lSJ}(\epsilon ) - 2^{-5/2}\,,
\label{b2} 
\ee
where $\epsilon=p^2/2m$ is the kinetic energy and the sum is over allowed partial waves.
The number density $n$ and the energy density $\E$ are calculated  in terms of the $b_2$ 
coefficient and are given by~\cite{Horowitz2006}
\begin{eqnarray}
n &=& \frac{2}{\lambda^3} (z_n+2z^2_n b_2) \,, \nonumber \\
\E &=& \frac{3T}{\lambda^3}\left [z_n+z_n^2\left (b_2-\frac{2}{3}Tb_2'\right )\right ] \,,
\end{eqnarray}
where $b_2' = db_2/dT$. The respective expressions for the HF calculation in pure neutron matter
are
\begin{equation}
\begin{split}
n &= 2 \int \frac{dk^3}{(2\pi)^3} f(\xi(\vec{k})) \,, \\
\E &= 2  \int \frac{dk^3}{(2\pi)^3} \left ( \varepsilon(\vec k) - \frac{1}{2} 
\Sigma_n(\vec{k}) \right ) f(\xi(\vec{k}) ) \,.
\end{split}
\end{equation}

\begin{figure}[t]
\centering
\includegraphics[scale=0.45]{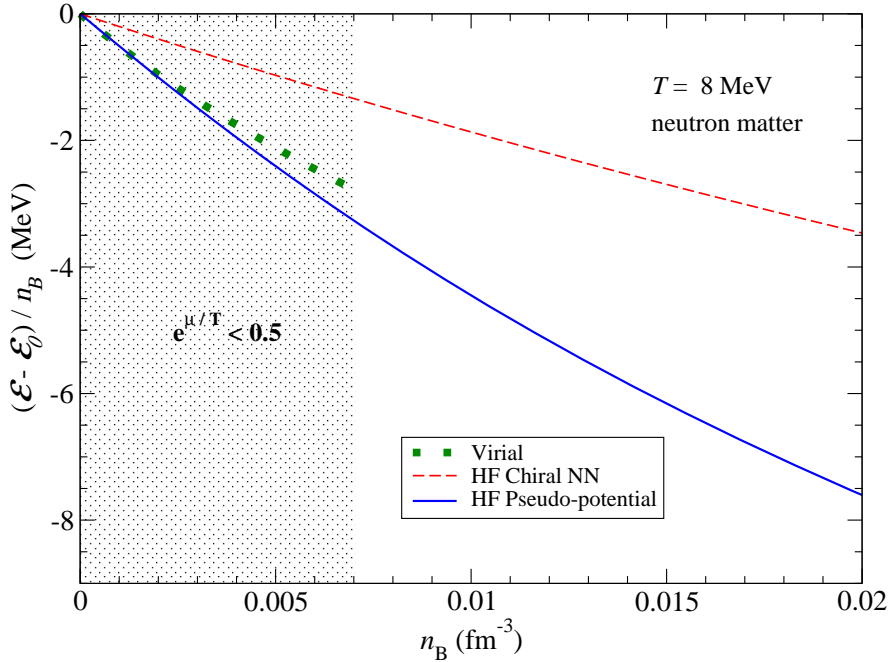}
\caption{(Color online) Change in the energy per particle of neutron
matter from NN interactions in the Hartree-Fock (HF) approximation. Results for both
the chiral NN potential and the pseudo-potential are shown and compared to the model-independent
virial equation of state~\cite{Horowitz2006}. The shaded area denotes
the density region in which the fugacity $z < 0.5$.}
\label{fig:virialhf}
\end{figure}

A detailed study of low-density hot matter in the virial expansion is
presented in Refs.~\cite{Horowitz:2005nd,Horowitz2006}. Here we
consider neutron matter and use the second virial coefficient
computed in Ref.~\cite{Horowitz2006} to compare with the results obtained 
using the chiral NN potential and the pseudo-potential in the HF approximation. 
Results for $T=8$\,MeV are displayed in
Fig.~\ref{fig:virialhf}, which shows the change in the energy per
particle due to NN interactions. At very low
densities (with corresponding fugacities $z < 0.25$), the virial
equation of state is well reproduced at the HF level when the
pseudo-potential is used, in agreement with previous statistical-mechanics
consistency checks~\cite{Fukuda56,Riesenfeld56}. At the breakdown
scale of the virial expansion $e^{\mu/T} \sim 0.5$, the
pseudo-potential predicts additional attraction over the virial
equation of state due to using full Fermi-Dirac distribution
functions. On the other hand, the  chiral
NN potential when used in the HF approximation significantly underestimates the strength of the
attractive mean field at low densities and therefore provides a
conservative upper bound on the energy per particle at temperatures
and densities beyond the scope of the virial expansion. Higher-order
perturbative contributions from chiral NN interactions are 
attractive and could 
lead to a narrower uncertainty band for the energy per particle.
We omit contributions from three-neutron forces, which are small
at these low densities.

A comparison between the second-order virial calculation and the HF calculation of matter 
with a finite proton fraction $Y_p = n_p / (n_p + n_n)$, where $n_n$ and $n_p$ are the neutron and 
proton densities, is complicated by the presence of the deuteron bound state. The HF description solely in terms 
of neutrons and protons will fail at low temperature and density when there is a large abundance of deuterons and light nuclei. 
However, on general grounds we expect the abundance of weakly bound states such as the deuteron to decrease rapidly 
with increasing temperature and density. The second-order virial calculation provides a correct description of deuterons at 
low density and moderate temperature, but it does not capture the physics relevant to the dissolution of weakly bound states 
with increasing density. Finite-density effects due to Pauli blocking of intermediate states in the $T$-matrix and modifications to the 
nucleon  propagators alter the scattering in the medium at low momentum. Recent calculations have shown that this leads to a 
decrease in the binding energy of light nuclei \cite{Ropke:2009}. These results indicate that the deuteron abundance is suppressed 
for $n_B \gtrsim 0.005$ fm$^{-3}$ \cite{Ropke:2009,Typel:2010,Hempel:2011kh} even at relatively low temperatures. Since the 
typical densities encountered in the neutrino-sphere are larger, especially during the proto-neutron star phase, in the following we 
will neglect the deuteron pole and calculate the nucleon self-energies in the HF approximation using both the chiral NN 
potential and the pseudo-potential. In Appendix \ref{appendix_a} we present a brief assessment of the deuteron contribution to the 
second-virial coefficient to show that it is relatively small at the densities and temperatures of interest.

\begin{figure}[h]
\centering
\includegraphics[scale=0.45]{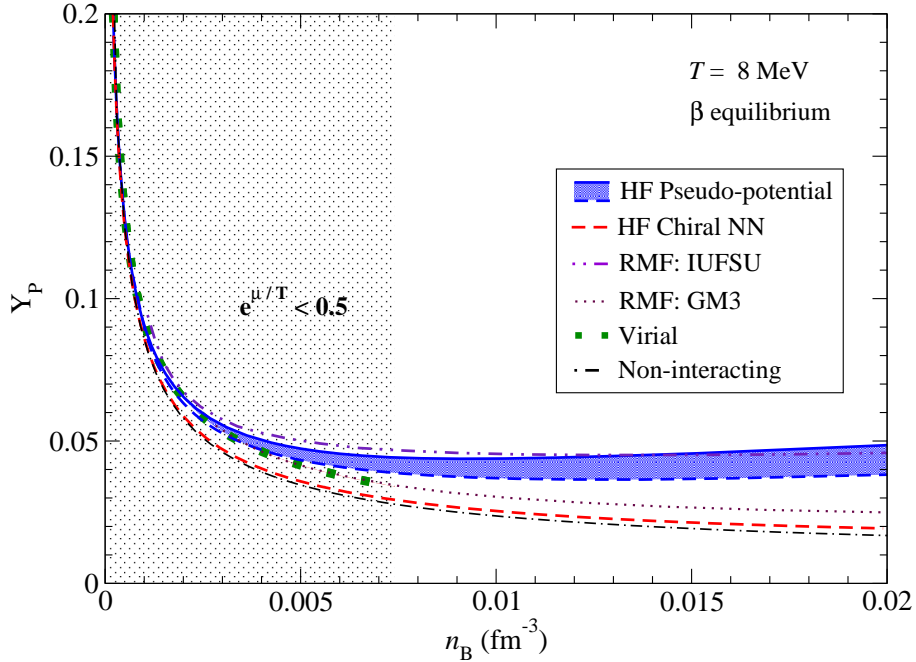}
\caption{(Color online) The proton fraction $Y_p$ as a function of density for matter in beta-equilibrium at temperature $T=8$\,MeV. 
Results for the chiral NN potential and the pseudo-potential in the Hartree-Fock (HF)
approximation are shown. The shaded $Y_p$ band is enclosed by solid and dashed lines resulting from the 
pseudo-potential and modified pseudo-potential calculations, respectively. The region
beteen the HF chiral and HF pseudo-potential band should be considered as a conservative uncertainty range. In addition, we compare to the model-independent 
virial equation of state \cite{Horowitz2006} as well as the predictions from relativistic mean-field 
(RMF) theory \cite{Roberts12c}. The shaded area denotes the density region in which the fugacity $z < 0.5$.}
\label{fig:yp_compare}
\end{figure}

To make a comparison between the HF and viral results for hot matter containing protons we consider 
neutron-rich matter at temperature $T = 8$\,MeV and 
determine the proton fraction in charge-neutral matter in beta-equilibrium for baryon densities in the range $n_B = 0.0001 - 0.02$\,fm$^{-3}$. We solve 
for the proton and neutron single-particle energies self-consistently and use them to obtain the proton and neutron densities given by
\begin{equation}
n_i = \frac{1}{\pi^2} \int_0^\infty p^2 dp \frac{1}{e^{(p^2/2M_i + \Sigma_i(p) - \mu_i)/T}+1} \,.
\end{equation} 
Attractive interactions between neutrons and protons increase the proton fraction $Y_p$ relative to the non-interacting case as is evident
from Fig.~\ref{fig:yp_compare}, which shows the proton fraction as a function of the density from different treatments of nuclear interactions. At the 
lower densities where the virial expansion is reliable, the HF pseudo-potential matches its predictions well. The HF calculation
with the chiral potential underestimates the attraction between neutrons and protons and predicts lower values of $Y_p$. 

Since the HF calculation does not provide a reliable treatment of the  deuteron pole in the neutron-proton $^3S_1$ channel, which is nonetheless included in defining the
pseudo-potential, we study how the results are affected when we modify the low-energy $^3S_1$ phase shifts. The alteration is designed to replace the bound state by a scattering resonance at low momentum and to asymptotically match with the experimental values of the phase shifts at high momenta. Further details can be found in Appendix~\ref{appendix_a}. By using the original and altered phase shifts in this channel we are able to provide a theoretical band for the prediction of the HF pseudo-potential approach as shown in Fig.\ \ref{fig:yp_compare} and in all future plots where the pseudo-potential results are shown.

\begin{figure}[t]
\centering
\includegraphics[scale=0.49]{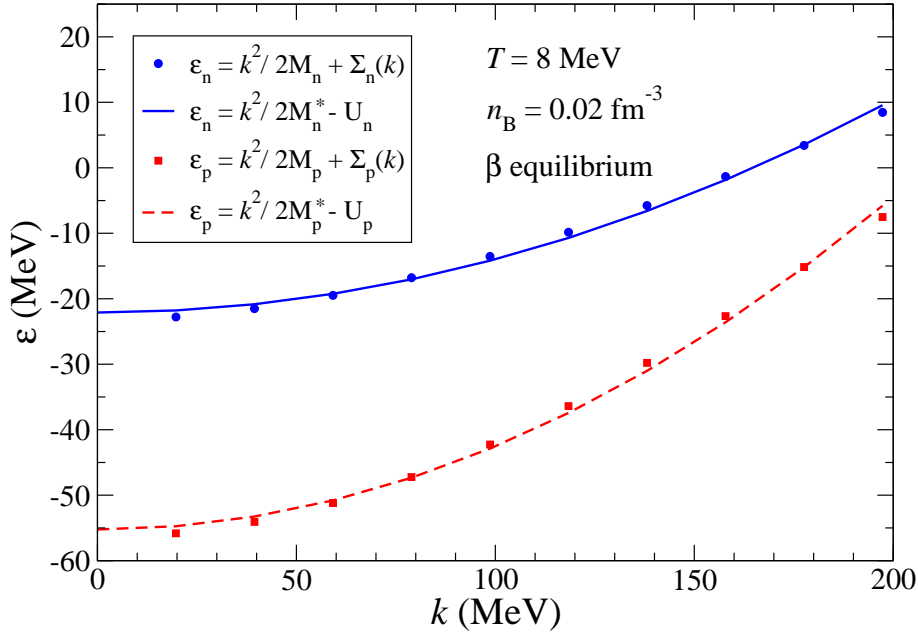}
\caption{(Color online) Momentum dependence of the neutron and proton
single-particle energies in hot ($T=8$\ MeV) and dense ($n_B =
0.02$~fm$^{-3}$) beta-equilibrated nuclear matter calculated in the HF approximation from the
pseudo-potential. The solid and dashed
lines are parametrized fits, with the form given in Eq.~(\ref{effmass}), of the non-relativistic 
dispersion relations for protons and
neutrons respectively.}
\label{fig:fitting}
\end{figure}

The ambient conditions encountered in the neutrino-sphere span densities and temperatures in the range 
$n_B =0.001-0.05$\,fm$^{-3}$ and $T=3-8$ MeV.  To study the nuclear medium effects, we choose baryon density $n_B =
0.02$\,fm$^{-3}$ and temperature $T = 8$\,MeV to compare with earlier results obtained  in Ref.~\cite{Roberts12c}. 
For these conditions the pseudo-potential predicts a proton fraction of 
$Y_p = 0.049$ (modified pseudo-potential: $Y_p = 0.038$), while for the HF chiral NN potential we find $Y_p = 0.019$. 
The neutron and proton momentum-dependent single-particle energies 
associated with mean-field effects from the nuclear pseudo-potential
are shown with filled circles and squares in Fig.~\ref{fig:fitting}, and 
qualitatively similar results were found for the chiral NN potential and modified pseudo-potential.
For convenience in calculating the charged-current reaction rates described later
in the text, we parametrize the momentum dependence of the single-particle energies 
with an effective mass plus energy shift:
\begin{equation}
\varepsilon(k) = \frac{k^2}{2M} + \Sigma(k) \simeq \frac{k^2}{2M^*} - U \,,
\label{effmass}
\end{equation}
where $U$ is momentum independent. To demonstrate that the quadratic form in Eq.~(\ref{effmass}) provides a good 
description, we display in Fig.~\ref{fig:fitting} the single-particle energies computed for the 
pseudo-potential (points) and quadratic fit (curves). The results for the proton and neutron 
effective masses and energy shifts are presented in Table~\ref{tab:shifts}. The Hartree-Fock energy from the chiral NN potential
is considerably smaller for both protons and neutrons than those obtained using the
pseudo-potential. The pseudo-potential predictions are also higher than those obtained in the relativistic mean-field (RMF) models employed in recent astrophysical simulations \cite{Roberts12c,Martinez-Pinedo12}. Simple RMF models such as the GM3 model from Ref.~\cite{Glendenning91} provide a fair description of symmetric nuclei but fail to reproduce ab-initio neutron matter calculations and are therefore not suitable for asymmetric matter calculations. In contrast, a new class of RMF models, such as the IUFSU model from Ref.~\cite{Fattoyev10}, that are constructed to simultaneously provide a good description of nuclear masses, neutron skin measurements, and match ab-initio calculations of pure neutron matter predict larger energy shifts and are closer in magnitude to those obtained using the HF pseudo-potential approach.  
 
\vspace{.2in}
\begin{table}[htbp]
\centering
\begin{tabular}{@{} ccccccc @{}} 
\hline 
Model & $Y_p $ & $M^*_n/M_n\ $ & $M^*_p/M_p\ $ & $U_n\ $ & $U_p\ $ & $\Delta U\ $ \\
\hline 
HF Pseudo-potential & 4.9\% & 0.65 & 0.42 & 22 & 55 & 33\\
HF Pseudo-potential (mod)\,\, & 3.8\% & 0.78 & 0.57 & 18 & 42 & 23\\
HF Chiral NN & 1.9\%  & 0.94 & 0.90 & \hspace{.07in}7 & 10 & \hspace{.07in}3  \\
RMF: GM3 & 2.5\% & 0.96 & 0.96 & 14 & 23 & \hspace{.07in}9 \\
RMF: IUFSU & 4.0\% & 0.94  & 0.94 & 31& 52 &21 \\
RMF: DD2 & 4.2\% & 0.92 & 0.92 & 9 & 25 & 16 \\
\hline
\end{tabular}
\caption{The Hartree-Fock (HF) effective masses $M^*$ and energy shifts $U$ (in units of MeV) for 
protons and neutrons in beta equilibrium
at $n_B = 0.02$\,fm$^{-3}$ and temperature $T = 8$\,MeV.
The difference in proton and neutron mean-field shifts is given by 
$\Delta U = U_p - U_n$, and the proton fraction is denoted by $Y_p$. 
Results for both the pseudo-potential and its modified (mod) version are compared to those from the chiral NN interaction
and RMF models \cite{Roberts12c,Martinez2014}. }

\label{tab:shifts}
\end{table}

\begin{figure}[h]
\centering
\includegraphics[scale=0.49]{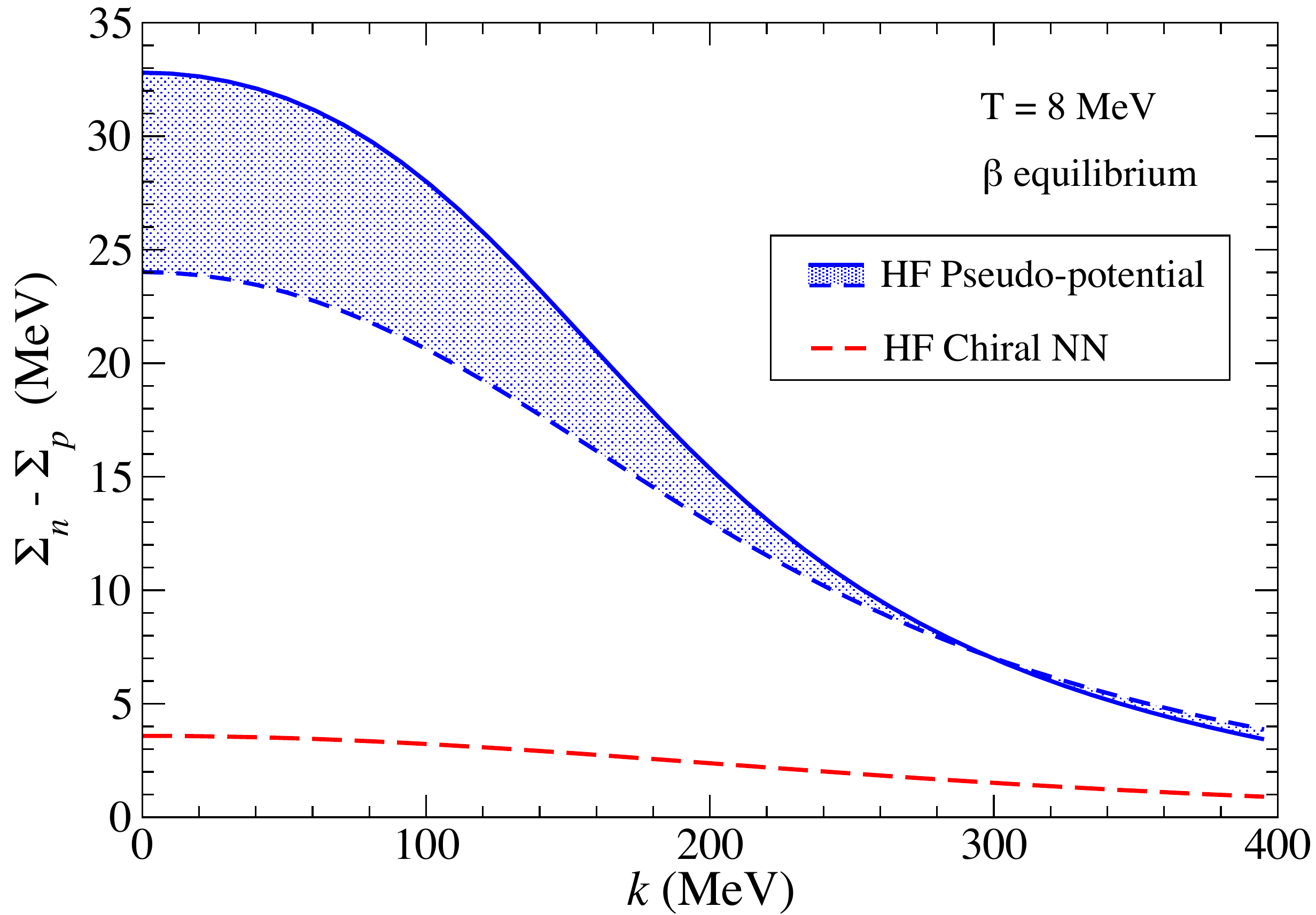}
\caption{(Color online) Difference in the momentum-dependent self-energies of 
neutrons and protons in the HF
approximation for beta-equilibrated matter
at $n_B = 0.02$\,fm$^{-3}$ and temperature $T = 8$\,MeV. Results for
the chiral NN potential and pseudo-potential are shown.}
\label{fig:sedif}
\end{figure}

\begin{figure}[h]
\centering
\includegraphics[scale=0.48]{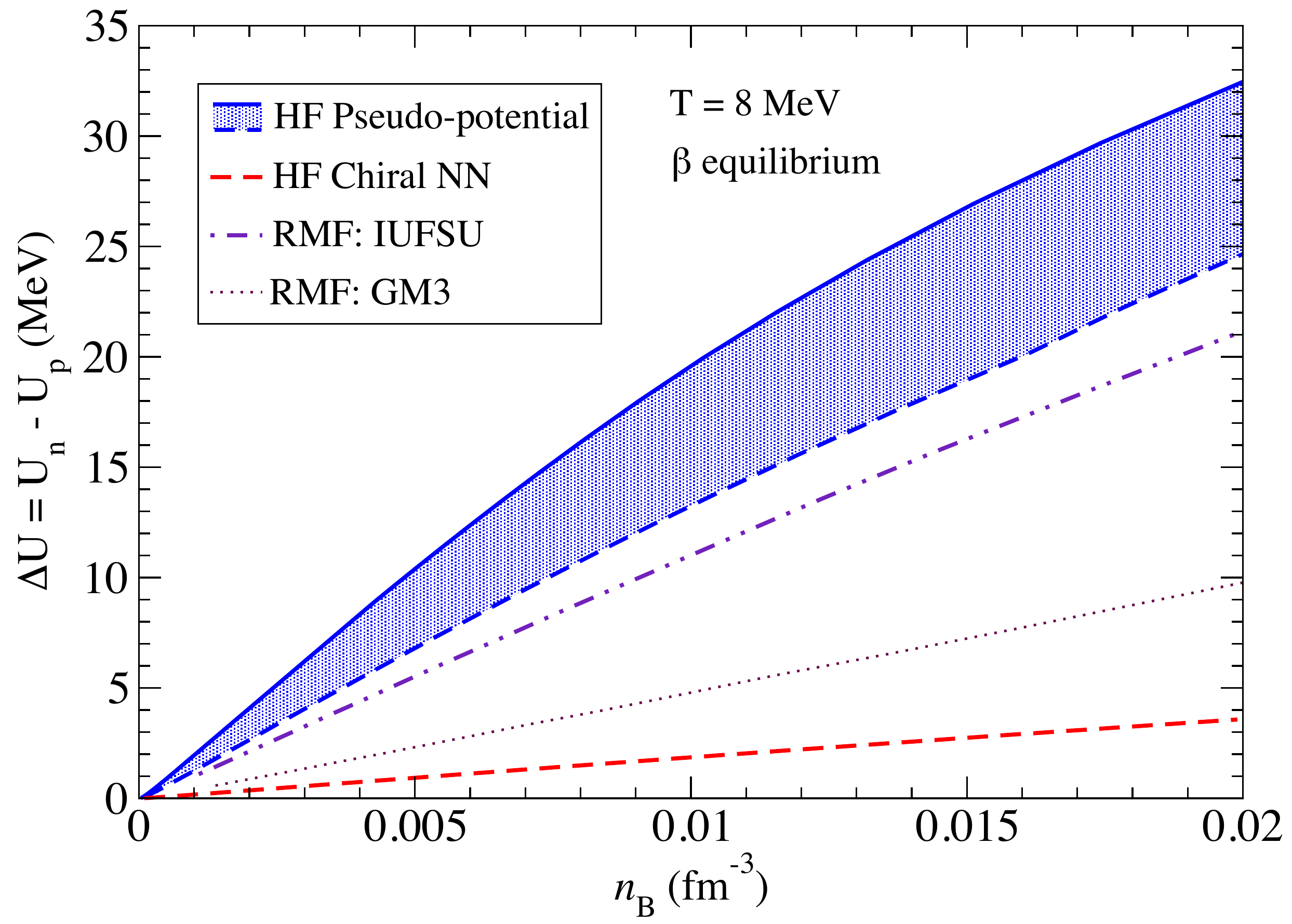}
\caption{(Color online) Difference in the HF proton and neutron energy shifts $U$, 
defined in Eq.\ (\ref{effmass}), as a function of the density. The results from the pseudo-potential and 
chiral NN interaction are compared to those from RMF models \cite{Roberts12c}. }
\label{fig:duplot}
\includegraphics[scale=0.48]{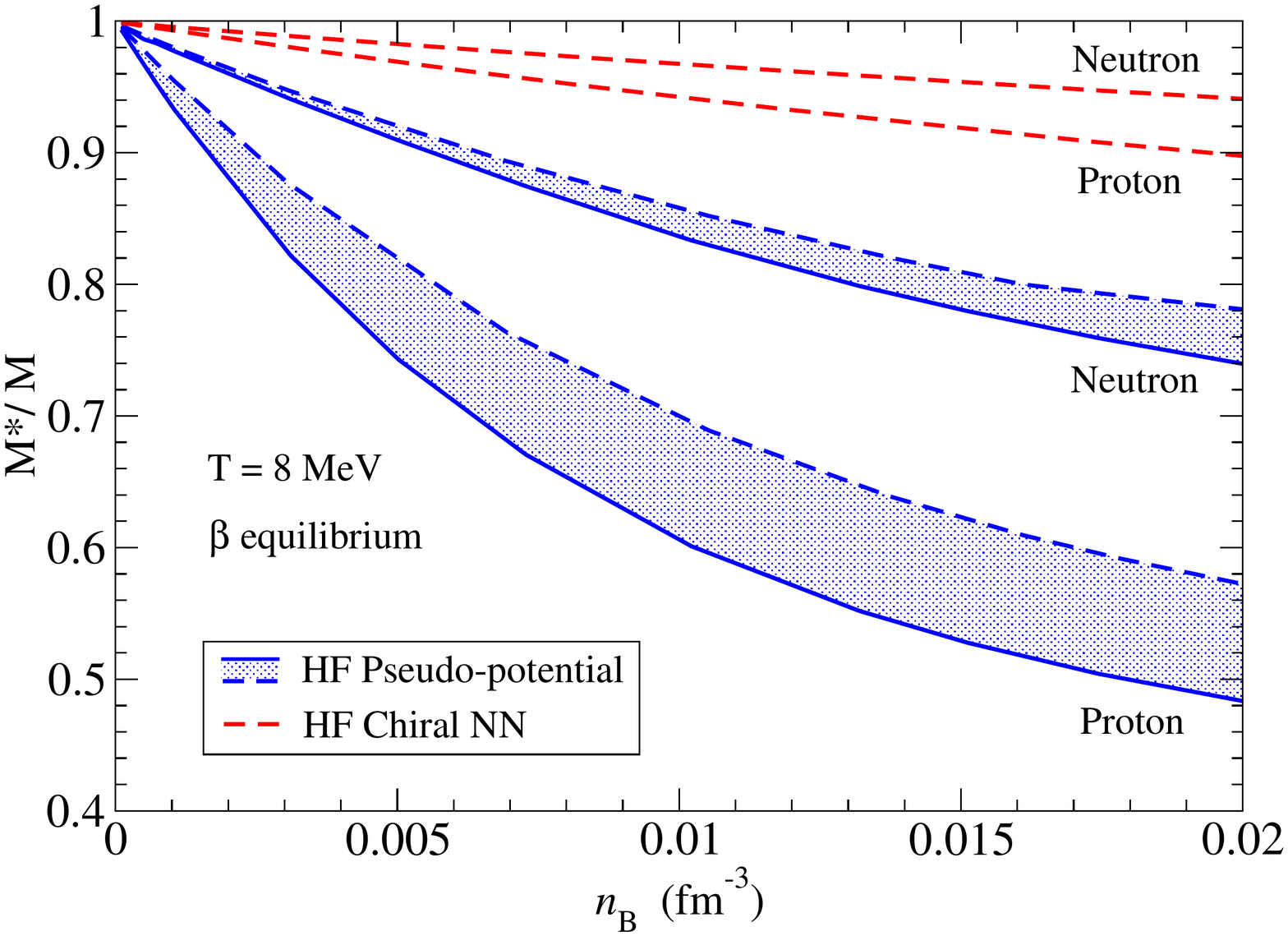}
 \caption{(Color online) Proton and neutron effective masses $M^*/M$ in the HF approximation 
 as a function of the density for the pseudo-potential and chiral NN interaction. }
 \label{fig:msplot}
\end{figure}

In Fig.~\ref{fig:sedif} we show the difference in the neutron and
proton self-energies for the chiral NN potential and the
pseudo-potential. The momentum dependence is also quite different for these
two cases. While the effective masses of proton and neutron quasiparticles
are similar and close to bare masses when chiral NN interactions are treated in the HF
approximation, the implicit iteration of NN interactions in
the pseudo-potential results in proton and neutron effective masses that are quite different from each other
and much smaller than the free-space masses. The density dependence  of self-energy shifts and nucleon effective masses are shown in Figures~\ref{fig:duplot} and~\ref{fig:msplot} respectively. As discussed earlier, the band for the pseudo-potential represents the variation expected for different treatments of the low-momentum behavior of the potential in the deuteron channel.  
The rapid increase in $\Delta U=U_n-U_p$ and a similarly rapid decrease in the nucleon effective masses obtained in the HF pseudo-potential approach are quite intriguing. Although the HF pseudo-potential approach is well motivated at these low densities and high temperatures as discussed earlier, these predictions for the self-energies are surprisingly large and have to be tested with explicit higher-order calculations in the many-body expansion. For now, it would be reasonable to suppose that the range spanned by the predictions of the HF chiral and HF pseudo-potential approaches represents our current uncertainty associated with non-perturbative effects. Using this as a representative range we discuss in the following section how the energy shifts and effective masses influence the neutrino/antineutrino  mean free paths  at the temperatures and densities of relevance to the neutrino-sphere.
 

\section{Neutrino absorption mean free path}
\label{sec:WeakRates}

\begin{figure}[h]
\centering
\includegraphics[scale=0.5]{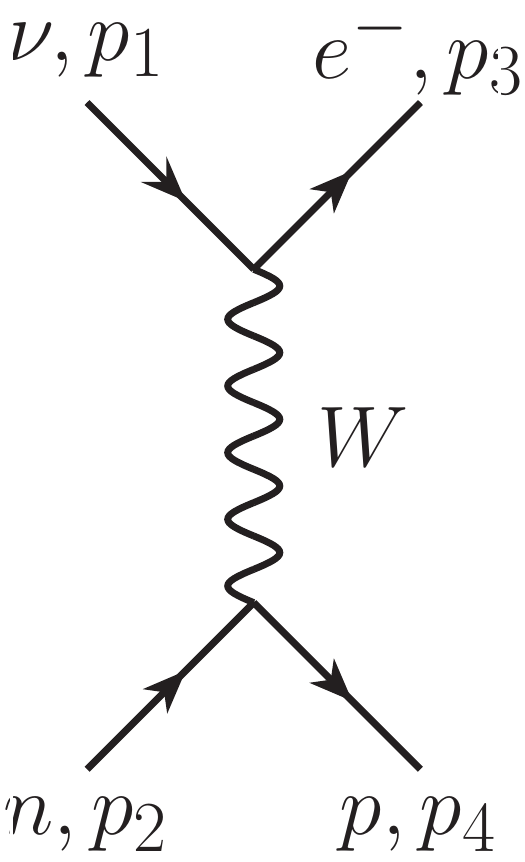}
\caption{Tree-level scattering amplitude for the process 
$\nu_e+n \rightarrow e^-+p$\,.}
\label{fig:weak_tree}
\end{figure}

The differential cross-section for the reaction $ \nu_e + n
\rightarrow e^{-} + p$ follows from Fermi's golden rule and is given
by
\begin{equation}
 \frac{d\sigma}{V} = \frac{2}{(2\pi)^5}\int d^3p_n\, d^3p_e\, d^3p_p~ \mathcal{W}_{fi}~{\delta}^{(4)}(p_{{\nu}_e} +p_n - p_e - p_p)  f_n(\xi_n) (1-f_e(\xi_e)) (1-f_p(\xi_p)) \,,
\label{dsigma}
\end{equation}
where $f(\xi)$ and $E$ are the fermi distribution functions and energies of the particles, and 
\begin{equation}
\mathcal{W}_{fi}=\frac{\langle |\mathcal{M}|^2 \rangle}{2^4 E_n E_p E_e  E_{{\nu}_e}}
\label{eq:matrix}
\end{equation}
is the transition probability. $\langle|M|^2\rangle$ is the
squared matrix element (corresponding to the diagram in
Fig.~\ref{fig:weak_tree}), averaged over initial spin states and
summed over the final spin states. For the reaction 
$\bar{\nu}_e + p \rightarrow e^{+} + n$ one obtains a similar
expression but with the replacement: $n \leftrightarrow p$, $e^-\rightarrow e^+$, and
$\nu_e \rightarrow \bar{\nu}_e$.  To simplify notation, we label the incoming
neutrino as particle $1$ with four-momentum $p_1=(E_1,\vec{p}_1)$, the
incoming baryon as particle $2$ with four-momentum
$p_2=(E_2,\vec{p}_2)$, and the outgoing lepton and baryon by the
particle labels $3$ and $4$, with four-momenta $p_3=(E_3,\vec{p}_3)$
and $p_4=(E_4,\vec{p}_4)$, respectively.

In the non-relativistic limit, retaining only terms independent of
the nucleon velocity, Eq.~(\ref{dsigma}) simplifies to
\begin{equation}
\frac{1}{V}
\frac{d^2{\sigma}}{d\cos\theta \, dE_3} = \frac{G_F^2 \cos^2 \theta_C}{4 \pi^2} \, |\vec{p}_3| \, E_3 \, (1-f_3(\xi_3))
 \Bigl[(1 + \cos\theta)   S_\tau(q_0,q)+ g_A^2(3-\cos\theta) S_{\sigma \tau}(q_0,q)\Bigr]\,,
\label{eq:dcross}
\end{equation}
where $g_A=1.26$ is the nucleon axial charge, $\theta$ is the angle
between the initial-state neutrino and the final-state lepton, and
$\theta_C$ is the Cabibbo angle. $S_\tau(q_0,q)$ and $S_{\sigma
\tau}(q_0,q)$ are the response functions associated with the Fermi
and Gamow-Teller operators, respectively~\cite{Roberts12c}. The
energy transfer to the nuclear medium is $q_0 = E_1-E_3$, and the
magnitude of the momentum transfer to the medium is $q^2 = E_1^2 +
E_3^2 - 2 E_1 E_3\cos\theta_{13}$, because for the leptons
$|\vec{p}_1|=E_1$ and $|\vec{p}_3|=E_3$.  In general, the response
functions $S_\tau(q_0,q)$ and $S_{\sigma \tau}(q_0,q)$ are different
because of isospin and spin-isospin dependent correlations in the
medium~\cite{Burrows99,Reddy99}. However, in the HF approximation
$S_\tau(q_0,q) =S_{\sigma \tau}(q_0,q) =S_{\rm F}(q_0,q)$ where
\begin{equation}
S_{\rm F}(q_0,q) = \frac{1}{2\pi^2} \int d^3p_2 \delta(q_0 + E_2 - E_4) f(E_2)
(1-f(E_4)) \,,
\label{eq:response}
\end{equation}
is the response function for a non-interacting Fermi gas, and follows
directly from Eq.~(\ref{dsigma}). The effects of interactions are
included in Eq.~(\ref{eq:response}) by using the HF self-energies for
neutrons and protons calculated in Sec.~\ref{sec:HFenergies}. We
use the quadratic form defined by Eq.~(\ref{effmass}):
\begin{equation}
E_2 = M_2 +\frac{p_2^2}{M_2^{*}} - U_2 \quad {\rm and} \quad
E_4 = M_4 +\frac{p_4^2}{M_4^{*}} - U_4 \,,
\label{eq:dispersion}
\end{equation}
where $M_2, M_4$ are the physical masses, $M_2^{*},M_4^{*}$ are the
effective masses, and $U_2, U_4$ are the momentum-independent
interaction-energy shifts of the initial- and final-state nucleon,
respectively. It is also straightforward to include in the nucleon
currents corrections due to weak magnetism of order $|\vec{p}|/M$. To do so,
we explicitly calculate the square of the matrix element appearing in
Eq.~(\ref{eq:matrix}) for the Fermi weak interaction Lagrangian
\begin{equation} 
\CL = \frac{G_F}{2\sqrt{2}} \, \bar{\psi}_4 (g_V \gamma_\mu 
+ i g_M \, \frac{\sigma_{\mu\nu}q^\nu }{ M} - g_A \gamma_\mu \gamma_5) 
\psi_2 \cdot \bar{e}_3(\gamma^\mu-\gamma^\mu\gamma^5) \nu_1 \,,
\label{eq:lweakmag}
\end{equation}
where the $\psi$'s are the nucleon spinors and $e$ and $\nu$ are the
final-state lepton and initial-state neutrino spinors. The nucleon
current has a vector component with $g_V = 1$, an axial-vector
component with $g_A=1.26$, and a Pauli component that incorporates
weak magnetism with $g_M=3.71$~\cite{Horowitz02}. We find that the 
differential cross-section per unit volume can be written as
\begin{equation}
\frac{d \sigma(E_1)}{V \, d\Omega dE_3} = E^2_3 \, (1-f_3(E_3))
\int \frac{d^3 p_2}{(2\pi)^3} \, \mathcal{W}_{\rm fi} \, 
\delta (E_1+E_2-E_3-E_4) f_2(E_2)(1-f_4(E_4)) \,.
\label{eq:fulldcross}
\end{equation}
An explicit form for $\mathcal{W}_{fi}$ including weak magnetism and
leading $|\vec{p}|/M$ terms in the nucleon weak currents is derived in
Appendix~\ref{appendix_b}.

\begin{figure}[h]
\centering
 \includegraphics[scale=.55]{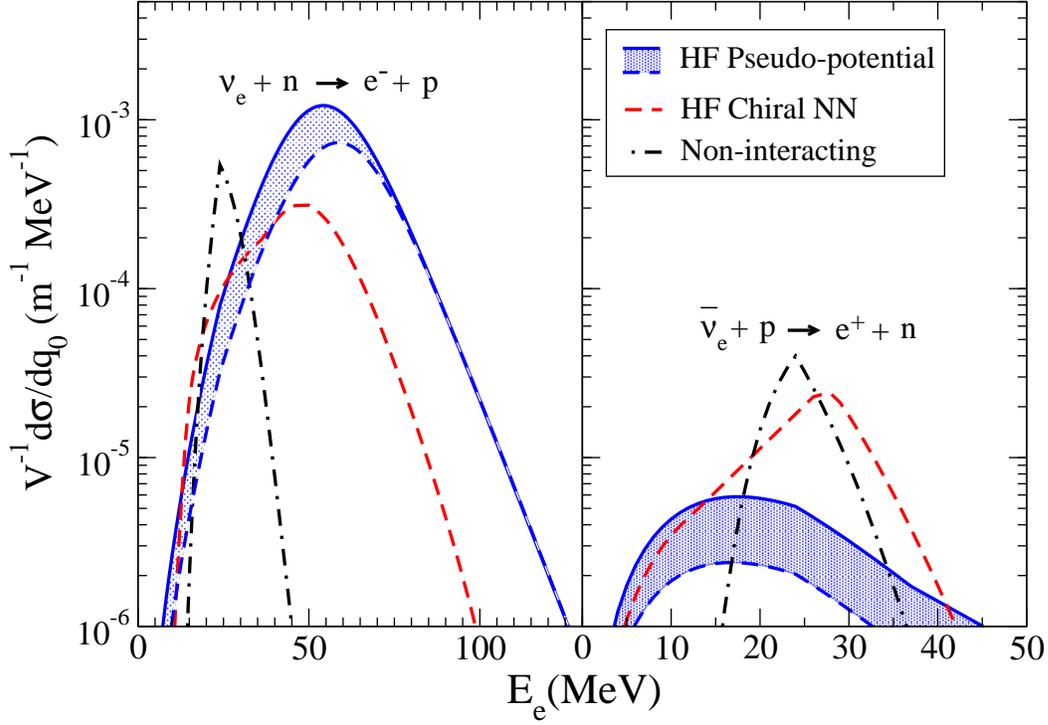}
\caption{(Color online) Effect of the in-medium neutron (proton) dispersion 
relation on the differential cross-section for (anti-)neutrino absorption 
 as a function of the outgoing lepton energy $E_e$. We consider an incoming neutrino energy  
 $E_\nu=24$ MeV and matter in beta equilibrium at a density $n_B=0.02$ fm$^{-3}$ and temperature 
$T=8$ MeV, including weak magnetism and leading $|\vec{p}|/M$ corrections. The
chiral NN interaction and pseudo-potential are both used in the HF 
approximation.  This provides a range for the theoretical uncertainty due to the many-body treatment, which
can be improved by performing higher-order calculations.}
\label{fig:diffmfp}
\end{figure}

The differential absorption rates for neutrinos and anti-neutrinos are
shown in Fig.~\ref{fig:diffmfp}.  The rates are shown as a function of
the energy of the outgoing lepton ($e^{-}$ for $\nu_e + n \rightarrow
p + e^{-}$ or $e^{+}$ for $\bar{\nu}_e + p \rightarrow n + e^{+}$) and
for an incoming neutrino energy of 24 MeV which is the mean thermal energy $E_\nu \sim 3 T = 24$ MeV at the ambient temperature of $T = 8 $ MeV. 
The trends seen in the figure can be understood on the basis of our 
earlier discussion of reaction kinematics in
Sec.~\ref{sec:Kinematics}, where it was shown that the interaction-energy
shifts in neutron-rich matter enhance the rate for $\nu_e$ absorption
and suppress the $\bar{\nu}_e$ rate. Further, since the energy shifts
are larger and the effective masses are smaller for the
pseudo-potential, charged-current rates calculated using the
pseudo-potential show larger differences than with the chiral NN
potential in the Born approximation.


\begin{figure}[h]
\centering
\includegraphics[scale=0.55]{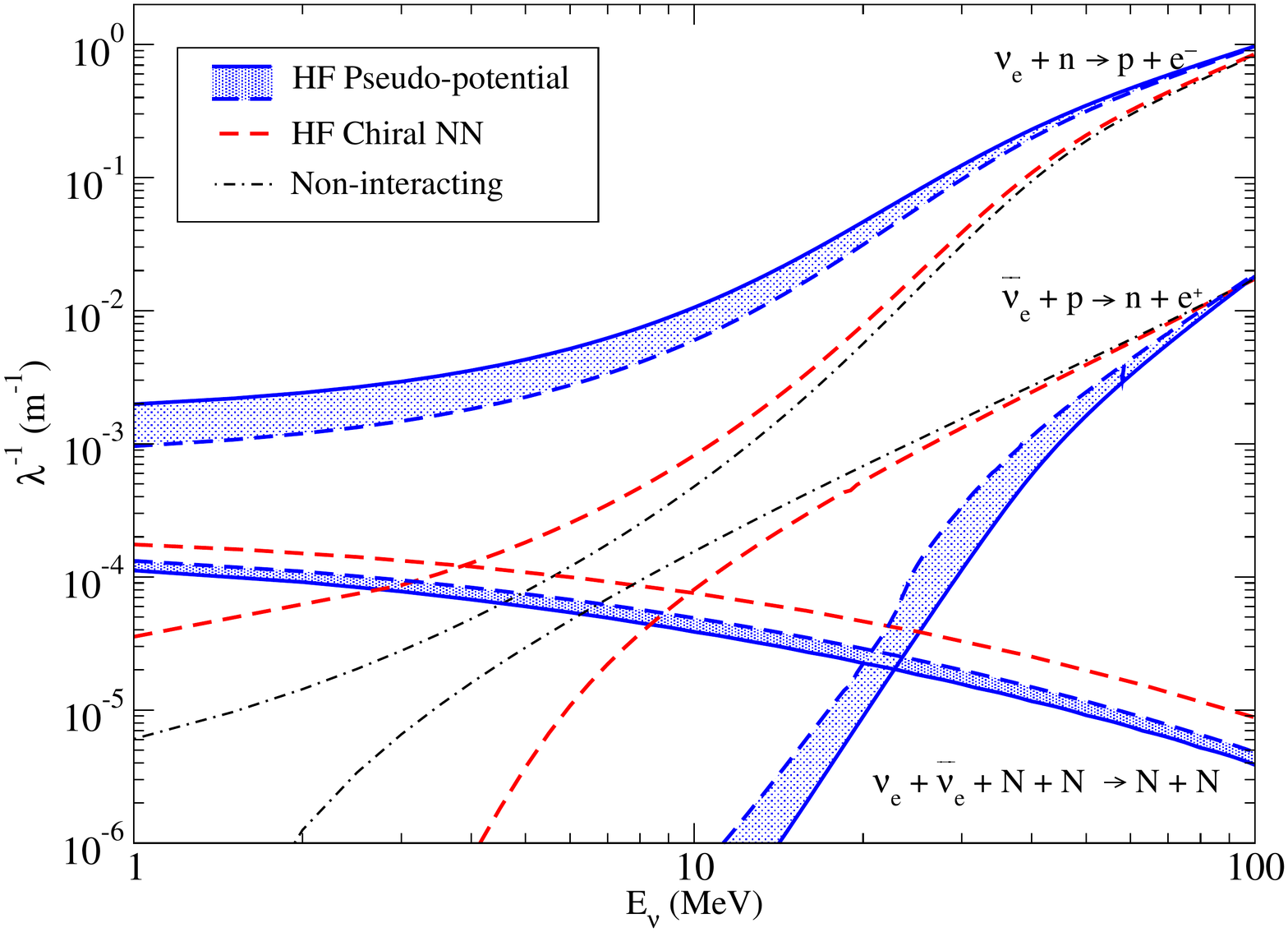}
\caption{(Color online) Effect of the in-medium neutron (proton) dispersion relation on the 
(anti-)neutrino absorption mean free path in beta-equilibrated matter at density 
$n_B=0.02$ fm$^{-3}$ and temperature $T=8$ MeV. The
chiral NN potential and pseudo-potential are both used in HF 
approximation. This provides a conservative range for the theoretical uncertainty due to the many-body treatment, which
can be improved by performing higher-order calculations.
Also shown is the mean free path for the neutrino-pair absorption process.}
\label{fig:mfptotal}
\end{figure}

The inverse neutrino mean free path for the absorption reactions mentioned, $ \lambda^{-1}_\nu(E_\nu)=
v_{\mathrm rel}\sigma/V$, where $v_{\mathrm rel}=c$ is the relative
velocity for relativistic neutrinos, can be calculated by numerical
integration of the differential cross-section defined in
Eq.~(\ref{eq:fulldcross}). The results shown in
Fig.~\ref{fig:mfptotal} follow the trends expected from the
results for the differential cross-section. The difference between
neutrino and anti-neutrino mean free paths is enhanced by the difference between 
the neutron and proton self-energies, and is larger for the case where the HF self-energy
was obtained using the pseudo-potential. Our range for the mean free
paths should be compared with those obtained in
Refs.~\cite{Martinez-Pinedo12,Roberts12c}. We refer to Fig.~1 in
Ref.~\cite{Martinez-Pinedo12} and Figs.~2 and~3 in
Ref.~\cite{Roberts12c} where similar results were obtained using a
phenomenological RMF model. Our results are
qualitatively similar to those obtained earlier, but important
quantitative differences exist. The $\nu_e + n \rightarrow e^- + p$ rate is enhanced by almost a factor of 7 relative
to the non-interacting case for $E_\nu=24$ MeV and the
$\bar{\nu}_e + p \rightarrow e^+ + n$ rate is suppressed by a larger factor $\simeq 30$.
 Under these conditions, neutral current scattering $\bar{\nu}_e + N \rightarrow \bar{\nu}_e + N$ 
 and the inverse bremsstrahlung process $\bar{\nu}_e + \nu_e + N +N \rightarrow N + N$, where $N$ can be either a neutron or a proton, can be expected to be more important.
 For energy exchange, the latter absorption process will be more relevant and is shown in  Fig.~\ref{fig:mfptotal}. Its rate is given by~\cite{Hannestad98}
\begin{equation}
\lambda_{\rm Brems}^{-1}(\omega_1)=2\pi\,G_F^2 \, n_B \int\frac{d^3\vec k_2}{(2\pi)^3}
\, (3-\cos\theta) f_2 S_A(\omega_1+\omega_2)\ ,
\end{equation}
where $\vec k_2$ is the momentum of the neutrino, $\omega_1$ and
$\omega_2$ are the energies of the anti-neutrino and neutrino,
respectively, $\theta$ is the scattering angle, and $f_2$ is the
occupation number of the neutrinos. We use the axial response function
$S_A(\omega)$ from Ref.~\cite{Bartl:2014hoa} and assume a
Maxwell-Boltzmann distribution at temperature $T$ for the neutrinos. 
The inverse mean free path due to the neutrino-pair absorption obtained using the chiral NN potential
is shown in red, and results obtained using the full $T$-matrix potential (corresponding to our pseudo-potential for
the self-energy calculations) is shown in blue using consistently the electron fractions and effective
masses given in Table~\ref{tab:shifts}.
\section{Conclusions}
\label{sec:Conclusions}

In this study we have presented a calculation of the HF self-energy of
protons and neutrons in the hot neutron-rich matter encountered in the
neutrino-sphere of supernovae and used them to calculate the
charged-current neutrino and anti-neutrino mean free paths. The mean free paths were found to be quite sensitive to the nucleon dispersion
relation, especially to the difference in the energy shifts
experienced by neutrons and protons in hot and relatively low-density
neutron-rich matter.  The difference between the results obtained
using a chiral N$^3$LO potential and the pseudo-potential is large and indicates that 
non-perturbative effects in the particle-particle channel, which are approximately included 
in the pseudo-potential, are important. A desirable feature of the HF pseudo-potential approach 
is that it reproduces the predictions of the virial calculation for the energy shifts which are exact in the 
low-density, high-temperature limit. However, the reader should not assume
the results obtained by the pseudo-potential at the densities and temperatures comparable to the ones 
displayed in Figs.~\ref{fig:diffmfp} and \ref{fig:mfptotal} to be the definitive answer. Instead, it would be prudent to 
treat the entire region between the HF chiral potential and pseudo-potential as a theoretical band, which needs to be further improved by higher-order 
many-body calculations. The uncertainty associated with not including the deuteron bound state contribution consistently in the HF pseudo-potential approach was studied by altering the low-momentum $^3{\rm S}_1$ phase shift to mimic the behavior expected from a low-energy resonance. This error was found to be relatively small in comparison at the relevant temperatures and densities.  Although the RMF model predictions are roughly consistent with the theoretical band it should be noted that they 
are constrained by fitting to the properties of nuclei, which are largely determined by the behavior at nuclear saturation density and
small isospin asymmetry and zero temperature. The error introduced by their extrapolation
to low density, large isospin asymmetry and high temperature can be large.  

At temperatures lower than those considered in the present work, the importance 
of Pauli blocking precludes the use of the pseudo-potential, and an alternative
strategy would be to employ an in-medium $T$-matrix as an effective interaction. This 
framework treats on equal footing quasiparticle energy shifts from the nuclear 
mean field and Pauli blocking in intermediate states, both of which tend to 
suppress the role of the strongly attractive components of the nuclear 
interaction. Consequently, we can expect that the large energy shifts reported in the present study 
should be reduced in this regime, while at higher temperatures the in-medium 
$T$-matrix and pseudo-potential results could be expected to match quantitatively.

The larger difference between neutrino and anti-neutrino rates
compared to the predictions of the RMF models will have an impact on
supernova nucleosynthesis. To quantitatively gauge its importance it
will be necessary to incorporate these new rates into supernova and
PNS simulations and predict the resulting neutrino
spectra. Qualitatively, we can anticipate a larger $\delta \epsilon$
that would favor smaller $Y_e $ in the neutrino-driven wind
compared to the predictions in Ref.~\cite{Roberts12c} based on the RMF models. 
Simulations that incorporate our current results will be able to ascertain if the change in 
$\delta \epsilon$ is large enough to favor conditions for a robust r-process in the standard 
supernova neutrino-driven wind scenario. In addition, our calculations of neutrino cross sections were
performed in the impulse approximation. Here we  
neglect vertex corrections (screening) and finite lifetime effects (damping), which arise because the weak
interaction amplitude involving different nucleons in the system will
interfere. 
These effects were studied within the purview of the RMF
model in Ref.~\cite{Roberts12c} and were found not to have as large of
an effect as the corrections due to energy shifts because the typical
energy and momentum transfer were large compared to characteristic
scales associated with temporal and spatial correlations,
respectively. Nonetheless, these effects, which are known to be
important in the study of neutral-current reactions, warrant further
investigation. They can be systematically studied using chiral EFT
interactions within self-consistent Green's functions where both particle-particle 
and particle-hole diagrams in the response function are partially re-summed. At 
the high densities and temperatures chosen for this study, alpha particles and light 
clusters are disfavored. However, for a better understanding of a wider range of 
ambient conditions encountered in the neutrino-sphere, the role of these microphysical 
effects will need to be investigated and incorporated in proto-neutron star simulations.  
We plan to explore these topics in future work.
    
\begin{acknowledgments}

We thank Kai Hebeler, Andreas Lohs, Luke Roberts, and Gang Shen for useful
correspondence, and George Bertsch, Charles Horowitz and Gabriel Mart{\'i}nez-Pinedo for useful discussions. 
The work of E.\ R.\ and S.\ R.\ was supported in part by grants from NUCLEI
SciDAC program and by the DOE Grant No.\ DE-FG02-00ER41132, J.~W.~H.~ acknowledges
support from DOE Grant No.\ DE-FG02-97ER41014, and the work of 
A.\ B.\ and A.\ S.\ was supported by a grant from BMBF ARCHES, the ERC Grant No.~307986
STRONGINT, the Helmholtz Alliance HA216/EMMI, and the Studienstiftung
des deutschen Volkes. A.\ S.\ thanks the Institute
for Nuclear Theory at the University of Washington for its hospitality
and the DOE for partial support during the completion of this work. This work was also facilitated through the use
of advanced computational, storage, and networking infrastructure
provided by the Hyak supercomputer system, supported in part by the
University of Washington eScience Institute.

\end{acknowledgments}

\bibliographystyle{apsrev4-1}
\bibliography{charged3_alex}

\clearpage
\appendix
\section{Deuteron contribution and the modified pseudo-potential}
\label{appendix_a}
\begin{figure}[h]
\centering
\includegraphics[scale=0.35]{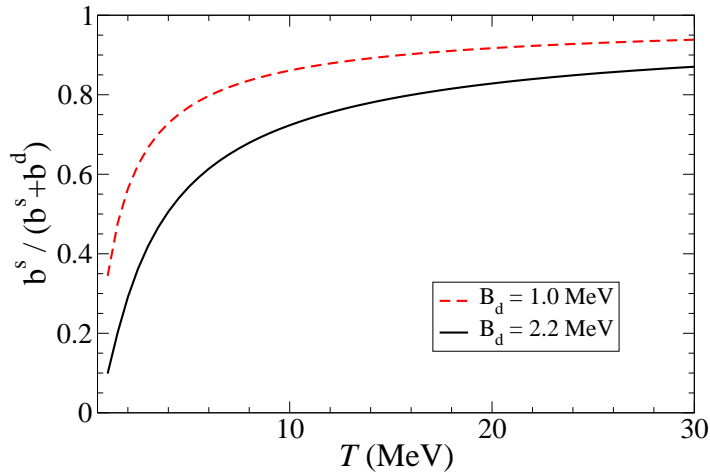}
\caption{(Color online) The relative importance of the scattering continuum contribution to the second virial coefficient (normalized with respect to the
sum of bound- and scattering-state contributions). Results assuming a free-space deuteron binding energy and a medium-reduced binding energy of 
$B_d = 1.0$\,MeV are shown.}
\label{fig:bnp}
\end{figure}

To assess the importance of the deuteron pole in neutron-proton scattering we study its contribution in the second-order virial calculation. 
Since the contributions to the second virial coefficient from the bound state denoted by $b^{\rm d}$ and the scattering continuum denoted by 
$b^{\rm s}$ can be calculated separately \cite{Horowitz:2005nd}, the ratio 
\begin{equation}
r=\frac{b^{\rm s}}{b^{\rm s}+b^{\rm d}}\,,
\end{equation}
is a measure of the relative importance of the scattering continuum. In the virial calculation $r$ is independent of density and increases rapidly 
with temperature as shown in Fig.~\ref{fig:bnp}. At the physical value of the deuteron binding energy the contribution from the 
scattering continuum is about $70$\% at $T=8$ MeV.  Medium effects mentioned earlier reduce the deuteron binding energy, and although such 
changes to $B_d$ are beyond the scope of the virial expansion, it is still useful to explore how the ratio $r$ changes for smaller values of $B_d$. 

The red-dashed curve in Fig.~\ref{fig:bnp} was obtained by setting $B_d=1$ MeV and shows how dramatically the deuteron contribution decreases 
with $B_d$. Model calculations of the reduction in the deuteron binding energy predict  $B_d < 1 $ MeV  for $n_B > 0.005$ fm$^{-3}$ for typical 
temperatures in the range $T=5-10$~MeV~\cite{Typel:2010}. 

\begin{figure}[h]
\centering
\includegraphics[scale=0.4]{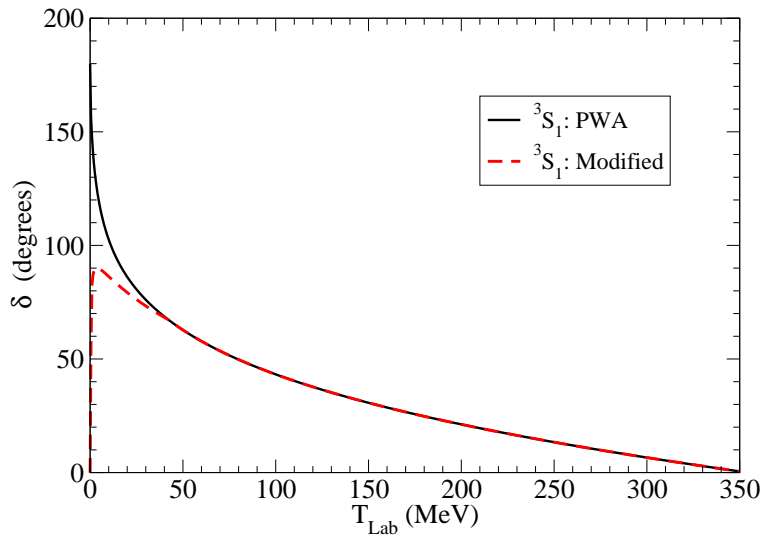}
\caption{(Color online) $^3S_1$ phase shift as a function of laboratory energy $T_{\rm lab}$ from the Nijmegen partial-wave analysis 
(PWA) \cite{Stoks93} as well as those used in the modified pseudo-potential. }
\label{fig:np3s1} 
\end{figure}

The neutron-proton scattering phase shift at low energies in the $^3S_1$ channel is dominated by the deuteron bound state and by Levinson's theorem is set equal to $\pi$ at zero momentum. This complicates the definition of the pseudo-potential that is to be used in the Born approximation since the potential constructed does not explicitly include these negative energy states. This in principle restricts the use of the pseudo-potential to large temperatures where we expect the deuteron abundances to be small. To assess the importance of the   
low-momentum behavior of the $^3S_1$  phase shifts we have modified them by hand. The modified potential mimics the low-momentum behavior expected for a resonance close to zero energy and asymptotically matches the original values of the $^3S_1$ phase shift at high momenta. We show both the original and modified versions of the $^3S_1$ phase shifts in Fig.~\ref{fig:np3s1}. 

\section{Transition rate including weak magnetism for non-relativistic nucleons}
\label{appendix_b}Here we derive expressions for the transition rate $\mathcal{W}_{fi}$ including the contribution from weak magnetism. We shall consider thermal neutrinos with energy $E_\nu \simeq T$ and calculate $\mathcal{W}_{fi}$  to linear order in  $T/M$ where $M$ in the  the average nucleon mass. The expressions are derived for neutrinos, and  analogous expression for anti-neutrinos are obtained by the substitution $\{g_V,g_M\} \rightarrow \{-g_V, -g_M \} $. 

The transition rate 
\begin{equation} 
\mathcal{W}_{fi}=\frac{\langle |\mathcal{M}|^2\rangle}{2^4 E_1 E_2 E_3 E_4}\,,
\end{equation}  
where 
\begin{equation}
 \begin{split}
 \langle|\mathcal{M}|^2\rangle & = \frac{1}{8} \ G_F^2 \hspace{0.1cm} {\rm Tr}\Bigl [\ \gamma^{\mu}(1-\gamma^5)\ \slashed{p}_1\ \gamma^{\nu}(1-\gamma^5)\ \slashed{p}_3\ \Bigr ] \\
& \times{\rm Tr} \biggl \{ \Bigl [\ \gamma_{\mu}(g_V - g_A \gamma^5)+i g_M\frac{\sigma_{\mu \alpha}q^{\alpha}}{ M}\ \Bigl  ](\slashed{p}_2 + M_2)\Bigl [\ \gamma_{\nu}(g_V - g_A \gamma^5)-i g_M\frac{\sigma_{\nu \alpha}q^{\alpha}}{ M}\ \Bigr ](\slashed{p}_4 + M_4)\biggl \} \\
& \equiv 8 \ G_F^2 ( \langle|\mathcal{M}|^2\rangle^{VA} + \langle|\mathcal{M}|^2\rangle^{VAM} +\langle|\mathcal{M}|^2\rangle^{M}) \,. 
\end{split}
 \end{equation}
is the square of the matrix element summed over final-state spins and averaged over initial-state spins for the interaction in Eq.~(\ref{eq:lweakmag}). Here, the vector-axial part is given by
 \begin{equation}
  \begin{split}
   \langle|\mathcal{M}|^2\rangle^{VA} = (g_A - g_V)^2\  &(p_1 \cdot p_4) (p_2 \cdot p_3) + (g_A + g_V)^2\ (p_1 \cdot p_2) (p_3 \cdot p_4) + (g_A^2 - g_V^2)\ M_2 M_4 \ (p_1 \cdot p_3)\ ,\\
  \end{split}
  \end{equation}
  the mixed term is given by
  \begin{equation}
   \begin{split}
    \langle|\mathcal{M}|^2\rangle^{VAM} =- \frac{g_M}{M}\biggl \{\ &(p_1 \cdot q)\Bigl[\ (2 g_A-g_V) M_4 (p_2 \cdot p_3)+(2 g_A+g_V)M_2 (p_3\cdot p_4)\ \Bigr]+(p_4\cdot q)g_V M_2 (p_1\cdot p_3)\\
    &-(p_2 \cdot q) g_V M_4 (p_1\cdot p_3) - (p_3\cdot q) \Bigl [\ 2 (g_A + g_V)M_4 (p_1 \cdot p_2)- (2 g_A-g_V) M_2 (p_1\cdot p_4)\ \Bigl ]\ \biggl \}\ ,
   \end{split} 
  \end{equation}
 and the contribution due to weak magnetism is given by 
  \begin{equation}
   \begin{split}
    \langle|\mathcal{M}|^2\rangle^{M} =\frac{g_M^2}{ M^2} \biggl \{ \ &(p_1 \cdot q) \Bigl[\ (p_2 \cdot q)(p_3\cdot p_4)-(M_2M_4+p_2\cdot p_4)(p_3\cdot q)+(p_2\cdot p_3)(p_4\cdot q)\Bigr]\\
 &+  ( p_3 \cdot q ) \Bigl[\ ( p_1 \cdot p_4 ) ( p_2\cdot q) +( p_1 \cdot p_2 )(p_4\cdot q)\ \Bigr]\\
  &- q^2 \Bigl[\ (p_1\cdot p_3) (M_2M_4 - p_2\cdot  p_4) +2 \bigl[(p_1 \cdot p_4)( p_2\cdot p_3)+(p_1\cdot p_2)(p_3\cdot p_4)\ \bigr]\ \Bigr] \biggr \} \,.
  \end{split}
 \end{equation}
 Setting $q^{\mu}= p_1^{\mu}-p_3^{\mu}$, we find that these results confirm equations~(11) and~(12) in Ref.~\cite{HorowitzLi}. 

At this stage we have only neglected the electron and neutrino masses because they are small compared to the thermal energies of the lepton $E_1 \simeq |\vec{p}_1| \sim T$ and  $E_3 \simeq |\vec{p}_3| \sim T$ .  In addition, for typical  ambient conditions we consider here, nucleons are non-relativistic and non-degenerate thus $|\vec{p}_2| \sim \sqrt{MT}$ and $|\vec{p}_4|\sim \sqrt{MT}$. Since the nucleon mass is large compared to the temperature it is useful to define the following expansion parameters 
\begin{equation}
\begin{split}
&\begin{pmatrix}& \{\chi_1,\ \chi_3,\ \chi_0\} \\ &\{v_2,\ v_4,\ \chi_q\} \end{pmatrix} \equiv \begin{pmatrix}&\{E_1/M,\ E_3/M,\ q_0/M\} \\ &\{|\vec{p}_2|/M,\ |\vec{p}_4|/M,\ |\vec{q}|/M \}\end{pmatrix} \sim \begin{pmatrix} &T/M \\ & \sqrt{T/M} \end{pmatrix}
\end{split}
\label{eq:small}
\end{equation}
where elements in the first row are parametrical of order $T/M$ and elements in the second row are order $\sqrt{T/M}$. 
Using energy-momentum conservation, $p_4^{\mu} = p_1^{\mu}+p_2^{\mu}-p_3^{\mu}$ and expanding to linear order in $T/M$ we find that 
\begin{align*}
 \mathcal{W}_{fi} \approx  \frac{G_F^2}{2} \times \biggl\{  \Bigl [&g_V^2  (1 + \eta_{13})+ g_A^2 (3 -\eta_{13})\Bigr ]& &\mathcal{O}(1)&\\
 +\Bigl [&2\ g_M g_A \chi_q\  (\eta_{1q}- \eta_{3q})-(g_A^2+g_V^2) v_2 (\eta_{12} + \eta_{23})\Bigr ]& &\mathcal{O}(\sqrt{\frac{T}{M}})&\\
 +\Bigl [&\frac{1}{2}\bigl[\ (g_A+g_V)^2 \chi_1-(g_A-g_V)^2 \chi_3\ \bigr ] (1-\eta_{13})+ (g_A^2+g_V^2) v_2^2 \eta_{12} \eta_{23}\\
 &+2\ g_M g_A {v_2} {\chi_q}\ (\eta_{12} \eta_{3q}-\eta_{1q} \eta_{23})+g_M^2{\chi_q}^2 (1-{\eta_{1q}} {\eta_{3q}})\Bigl ] \biggr \}& &\mathcal{O}(\frac{T}{M})&
\end{align*}
where the first line contains terms of $\mathcal{O}(1)$, the second line contain terms of $\mathcal{O}(\sqrt{T/M})$ and 
the third and fourth lines contains terms of $\mathcal{O}(T/M)$. The angles between three vectors $\vec{p}_i$ and $\vec{p}_j$ is denoted by $\eta_{ij}$ and the angle between $\vec{p}_i$ and $\vec{q}$ is denoted by $\eta_{iq}$.


\end{document}